\documentclass[12pt]{article}
\usepackage{graphicx}
\usepackage{authblk}

\newcommand\pubnumber{NuPhys2016-Patrizii}
\newcommand\pubdate{}

\def\bologna{INFN Bologna, Italy\\}
\def\support{\footnote{the OPERA Collaboration is listed in the last page}}

\def\Title#1{\begin{center} {\Large #1 } \end{center}}
\def\Autore#1{\begin{center}{ \sc #1} \end{center}}
\def\Address#1{\begin{center}{ \it #1} \end{center}}
\def\onbehalf#1{\begin{center}{ \sc #1} \end{center}}

\newcommand\pubblock{\rightline{\begin{tabular}{l} \pubnumber\\
         \pubdate  \end{tabular}}}

\newenvironment{Abstract}{\begin{quotation}  }{\end{quotation}}
\newenvironment{Presented}{\begin{quotation} \begin{center} 
             PRESENTED AT\end{center}\bigskip 
      \begin{center}\begin{large}}{\end{large}\end{center} \end{quotation}}

             \def\bibit{\vspace{-0.2cm}\bibitem}
             
\begin{document}
\begin{titlepage}
\pubblock
\vfill
\Title{Results from the OPERA Experiment}
\vfill
\Autore{Laura Patrizii}
\Address{\bologna}
\onbehalf{on behalf of the OPERA Collaboration\support}
\vfill
\begin{Abstract}
 
The OPERA experiment reached its main goal by proving the appearance of $\nu_{\tau}$   in the CNGS $\nu_{\mu} $  beam. 
A  sample of five  $\nu_{\tau}$ candidates  was collected allowing to reject the null hypothesis at $5.1 \sigma$. The estimation of $\Delta m^2_{23}$  in ``appearance mode''  has been obtained.
Updates  on the search for $\nu_{\mu} \to  \nu_e$ oscillations and on the search for sterile neutrino mixing in the $\nu_{\mu} \to  \nu_e$ and  $\nu_{\mu} \to  \nu_\tau$ channels are also reported.
\end{Abstract}
\vfill
\begin{Presented}
NuPhys2016, Prospects in Neutrino Physics\\
Barbican Centre, London, UK, December 12~--~14, 2016
\end{Presented}
\vfill
\end{titlepage}
\def\thefootnote{\fnsymbol{footnote}}
\setcounter{footnote}{0}
\section{Introduction}
The OPERA  experiment at the Gran Sasso Lab was exposed to the  CNGS $\nu_{\mu}$  beam, 730 km away from the beam source.

The CNGS was a conventional neutrino beam  optimised for $\nu_{\tau}$ appearance search. Unlike other neutrino beams 
designed to measure $\nu_{\mu}$ disappearance at the  atmospheric squared-mass splitting scale,
the mean energy (17 GeV) of the CNGS   was not tuned at the oscillation maximum which  for $L = 730$~km is at $E_{\nu} \sim 1.5$ GeV, i.e. below the $\tau$ production threshold.
The prompt $\nu_{\tau}$ contamination was negligible, $\mathcal{O}(10^{-6})$; the $\nu_e$ component was relatively small:  in terms of CC interactions, the $\nu_{e}$ and $\bar{\nu}_{e}$ contaminations were together $< 1\%$. 

The total exposure to the CNGS beam ($17.97 \times 10^{19}$ protons on target, PoT) resulted in 19 505 neutrino interactions in the OPERA target fiducial volume.

The OPERA detector was  made of two identical super modules (SMs) each consisting of a
target section made of lead/emulsion-film modules, of a scintillator
tracker detector, needed to pre-localize neutrino interactions within the target, and of a muon spectrometer.
The topology of neutrino interactions were recorded in emulsion cloud chamber detectors (ECC bricks) with submicrometric spatial resolution. Each brick was a stack of 56 1 mm thick lead plates, and 57 nuclear emulsion films with a 12.7~$\times$~10.2 cm$^2$ cross section, a thickness of $\sim$10 X$_0$ and a mass of 8.3 kg. 
In the bricks, the momenta of charged particles were measured by their multiple Coulomb scattering in the lead plates.
A changeable sheet (CS) doublet consisting of a pair of emulsion films was attached to the downstream face of each brick. The full OPERA target was segmented in about 150 000 bricks, arranged in each SM in 31 walls.
  Downstream of each target wall two orthogonal planes of electronic target trackers (TTs), made of 2.6 cm wide scintillator strips, recorded the position and deposited energy of charged particles. A spectrometer, consisting of iron core magnets instrumented with resistive plate chambers  and drift tubes was mounted downstream of each target module. The spectrometers are used to identify muons, determine their
charge, and measure their momentum with an accuracy of about 20\%. A detailed description of the OPERA detector
can be found in Ref.~\cite{det}.
 
\section{Data processing}
Neutrino events were classified either as 1$\mu$, i.e. events with  at least one track tagged as a muon, or as 0$\mu$ \cite{eledet}. 
A dedicated program reconstructs tracks in the electronic detectors and builds a 3D probability map for bricks to contain the neutrino vertex. 
 The CS films of the brick with the highest probability are developed and analysed with high-speed automatic optical microscopes~\cite{eledet}, searching for tracks compatible with the TT prediction. 
The tracks found in the CS doublet are extrapolated to the most downstream film of the brick and then followed upstream in the brick until  the stopping point (primary vertex). 
A  procedure is then applied to detect charged and neutral decay topologies, secondary interactions or photon conversions in the neighborhood 
of the primary vertex.
If a secondary vertex is found a full kinematical analysis 
is performed extending the scanned volume and following the tracks also in the downstream bricks. 
This analysis integrates the complementary information provided 
by emulsions and electronic detectors, making use of 
the angles measured in the emulsion films, the momenta determined by multiple Coulomb scattering measured in the brick, the momenta measured by the magnetic spectrometers, and the total energy deposited in the instrumented target acting as a calorimeter~\cite{eledet}. The energy of photons and electrons is also estimated using calorimetric techniques~\cite{nue}.
 The details of the event analysis procedure are described in Ref.~\cite{decaysearch}.
 
\section{Results on neutrino oscillations}
\subsection{$\nu_{\mu} \rightarrow \nu_{\tau}$}
\label{sec-oscillation}
Five events   out of all 0$\mu$ events and  $1\mu$ events with $p_\mu <$ 15 GeV/c fulfill the topological and kinematical cuts required for $\nu_{\tau}$ candidates~\cite{5Tau}.
In one of them the $\tau$ lepton undergoes a muonic decay~\cite{taumu}, one event is a $\tau \rightarrow 3h$ decay~\cite{tau3h}, and three events are $\tau \rightarrow 1h$ decays~\cite{tau1, tau4}. 

The numbers of expected signal and background events are estimated from the simulated CNGS  flux~\cite{cngs}. The expected detectable signal events in the 0$\mu$ events and $1\mu$ samples are obtained using the reconstruction efficiencies and the $\nu_{\tau}$ event rate in the flux normalised to the detected $\nu_{\mu}$ interactions. A similar normalisation procedure is also used in the background expectation. The details of the signal and background estimation are described in Ref.~\cite{tau3h}.
The expected numbers of $\nu_{\tau}$ events for each decay channel are computed assuming $\Delta m^2_{23} = 2.44 \times 10^{-3}$~eV$^2$~\cite{pdg} and maximal mixing (see Table~\ref{tab:final}). 
The total expected signal amounts to $2.64 \pm 0.53$ events. 
 The total systematic uncertainty on the expected signal is then set to 20\%~\cite{5Tau}.

The main sources of background in the search for $\nu_{\tau}$ appearance are charmed particle decays, hadronic interactions and large-angle muon scattering (LAS). 
 The uncertainties on the charm  and hadronic backgrounds are 20\%~\cite{decaysearch} and 30$\%$~\cite{5Tau}, respectively. 
A recent re-evaluation of the LAS background led to a significant reduction of its contribution ~\cite{las}. From this study it follows that the number of LAS
background events that satisfy the selection criteria amounts to [$1.2\pm 0.1$(stat.)$\pm0.6$(sys.)]$\times10^{-7}/\nu_\mu^{CC}$ interactions.
The estimated background events for the analysed data set with the corresponding uncertainties are listed  in Table~\ref{tab:final}.
The total expected background amounts to $0.25 \pm 0.05$ events. 

\begin{table}[t]
\begin{center}
\scriptsize
\vspace{-0.5cm}
\begin{tabular}{l|c c c c|c|c}
\hline
\hline

Channel & \multicolumn{4}{|c|}{Exp. Background} & Exp. Signal & Observed  \\
        & Charm & Hadronic re-int & LAS & Total &  &   \\
\hline
$\tau \rightarrow 1h$   & 0.017 $\pm$ 0.003 & 0.022 $\pm$ 0.006  & --                  & 0.04 $\pm$ 0.01   & 0.52 $\pm$ 0.10   & 3  \\
$\tau \rightarrow 3h$   & 0.17 $\pm$ 0.03   & 0.003 $\pm$ 0.001  & --                  & 0.17 $\pm$ 0.03   & 0.73 $\pm$ 0.14   & 1  \\
$\tau \rightarrow \mu$  & 0.004 $\pm$ 0.001 &  --                & 0.0002 $\pm$ 0.0001 & 0.004 $\pm$ 0.001 & 0.61 $\pm$ 0.12   & 1  \\
$\tau \rightarrow e$    & 0.03 $\pm$ 0.01   &  --                & --                  & 0.03 $\pm$ 0.01   & 0.78 $\pm$ 0.16   & 0  \\
\hline
Total                   & 0.22 $\pm$ 0.04   &  0.02 $\pm$ 0.01   & 0.0002 $\pm$ 0.0001 & 0.25 $\pm$ 0.05   & 2.64 $\pm$ 0.53   & 5  \\
\hline
\hline
\end{tabular}
\caption{\small Expected signal and background events in the analysed data set~\cite{5Tau}.}
\label{tab:final} 
\end{center}
\end{table}

The significance of the observed $\nu_{\tau}$ candidates is evaluated  as the probability that the background  can produce a fluctuation greater than or equal to the observed number of events. 
 Two test statistics are used, one based on the Fisher's method, the other one based on the profile likelihood ratio. 
Both methods exclude the background-only hypothesis with a significance of 5.1 $\sigma$~\cite{5Tau}. 
The observed number of $\nu_{\tau}$ candidates is also compatible with the expectations in the three neutrino oscillation framework. 
Based on the number of observed signal candidates   $\Delta m^2_{23}$   has been evaluated in ``appearance mode'' for the first time. 
Assuming full mixing the 90\% C.L. interval for $\Delta m^2_{23}$ is $[2.0, 5.0] \times 10^{-3}$~eV$^2$~\cite{5Tau}.
\subsection{$\nu_{\mu} \to \nu_{e}$}
\label{sec-sterile}

The possibility to efficiently disentangle electrons from photon conversion in the ECC bricks bases the search  for oscillations in the  $\nu_{\mu} \to \nu_{e}$ channel.  A dedicated procedure is applied to $0\mu$ events aiming at identifying ``shower hints" from  track multiplicity in the changeable sheet doublets. An additional scanning of volume extending  from the most downstream film  up to the interaction vertex is performed in order to  reconstruct electromagnetic showers. Events with a  shower initiated by a single track emerging from the primary vertex are classified as $\nu_{e}$ candidates.
A first  result corresponding to $5.3 \times 10^{19}$ pot was published in Ref.~\cite{nue}. The search  has been extended to the whole data set yielding 34 $\nu_{e}$ candidates. The expected number of $\nu_{e}$ CC interactions due to the intrinsic beam contamination is $37 \pm 5$.  Background events amount to $1.2\pm0.1$. They arise from misidentified $\pi_0$  in $\nu_{\mu}$ interactions without a reconstructed muon and $\nu_\tau$ CC interactions with $\tau$ decaying into an electron.
In the whole energy range $2.9\pm0.4$ oscillated $\nu_{e}$  CC events 
are expected assuming sin${^2}2\theta_{13}$~=~0.098, sin${^2}2\theta_{23}$~=~1, $\Delta m{^2}_{31}$~=~2.44$\times$10$^{-3}$~eV$^{2}$, $\delta_{CP}$~=~0, and neglecting matter effects.
In conclusion, the number of observed events is compatible with the 3-flavour oscillation model.

\subsection{Search for sterile neutrino mixing}
\label{sec-sterile}
 The results on $\nu_{\tau}$ appearance  have been interpreted in the context of the 3+1 neutrino model deriving limits on oscillations induced by a massive sterile neutrino. 
Exclusion regions are obtained in the $(\Delta m^2_{41}, \sin^2 2 \theta_{\mu \tau})$ parameter space. The limits on $\Delta m^2_{41}$ are extended up to $10^{-2}$~eV$^2$ for relatively large mixing, $\sin^2 2 \theta_{\mu \tau} > 0.5$. 
At large values of $\Delta m^2_{41}$  ($ >$~1~eV$^2$), marginalising over the $CP$-violating phase, values of the effective mixing parameter $\sin^2 2 \theta_{\mu \tau} > 0.119$ are excluded at 90\% C.L.~\cite{OPERASterile}.

In Ref.~\cite{nue} the number of $\nu_e$ candidates was compared to the expectation from an approximated two-state model parametrised in terms of two effective parameters, $\Delta m^2_{new}$ and $\theta_{new}$. 
The approximation is valid assuming $CP$ conservation, neglecting standard oscillations, treated as a background, and for large values of $\Delta m^2_{new}$ ($ > $ 0.1 $eV^2$). 
To optimise the sensitivity only events below 30 GeV were considered. Six  events were observed to be compared to an expectation of $9.4 \pm 1.3$~(syst.). 
For large $\Delta m^2_{new}$ values the 90$\%$ C.L. upper limit on $\sin^{2} 2\theta_{new}$ is at 7.2 $\times 10^{-3}$. This analysis is being updated in the 3+1 neutrino model using the whole $\nu_e$ data sample.

\section{Conclusions}
The OPERA experiment has discovered  $\nu_{\tau}$ appearance with a significance of $5.1~\sigma$ observing 5 $\nu_{\tau}$ candidates with a background of 0.25 events. 

The results on $\nu_{\mu} \rightarrow \nu_{\tau}$ search, compatible with the standard $3 \nu$ model, have been used to constrain the parameter space of oscillations induced by a massive sterile neutrino. 
Limits on the sterile neutrino mixing have also been derived 
in the $\nu_{\mu} \rightarrow \nu_{e}$ appearance channel.

In order to estimate oscillation parameters with reduced statistical uncertainty an analysis using a selection with released cuts and multivariate techniques is being performed. The unique
feature of the OPERA experiment to identify all neutrino flavours will allow a joint fit of all oscillation data.

\noindent{\footnotesize \textbf {The~OPERA~Collaboration:}~N.~Agafonova$^a$, A.~Aleksandrov$^b$, A.~Anokhina$^c$, S.~Aoki$^d$, A.~Ariga$^e$, T.~Ariga$^{e,e1}$, D.~Bender$^f$, A.~Bertolin$^g$, I.~Bodnarchuk$^h$, C.~Bozza$^i$
% ...

 \noindent{\scriptsize $^{a}$INR - Institute for Nuclear Research of the Russian Academy of Sciences, RUS-117312 Moscow, Russia; 
$^{b}$INFN Sezione di Napoli, 80125 Napoli, Italy; 
$^{c}$SINP MSU - Skobeltsyn Institute of Nuclear Physics, Lomonosov Moscow State University, RUS-119991 Moscow, Russia; 
$^{d}$Kobe University, J-657-8501 Kobe, Japan; 
$^{e}$Albert Einstein Center for Fundamental Physics, Laboratory for High Energy Physics (LHEP), University of Bern, CH-3012 Bern, Switzerland;  
$^{e1}$Faculty of Arts and Science, Kyushu University, J-819-0395 Fukuoka, Japan;
$^{f}$METU - Middle East Technical University, TR-06800 Ankara, Turkey; 
% ...
$^{g}$ INFN Sezione di Padova, I-35131 Padova, Italy; 
$^{h}$JINR - Joint Institute for Nuclear Research, RUS-141980 Dubna, Russia; 
$^{i}$Dipartimento di Fisica dell'Universit\`a di Salerno and ``Gruppo Collegato'' INFN, I-84084 Fisciano (Salerno), Italy; 
% ...
}
\end{document}